\documentclass[aps, prl, twocolumn, superscriptaddress, showpacs ]{revtex4}
\usepackage{bm}
\usepackage[dvips]{graphicx}
\usepackage{dcolumn}
\usepackage{longtable}

\begin{document}

\title{Incidence of the Tomonaga-Luttinger liquid state on the NMR spin lattice relaxation in Carbon Nanotubes}
\author{Y.~Ihara}
\altaffiliation{ihara@lps.u-psud.fr}
\author{P.~Wzietek}
\author{H.~Alloul}
\affiliation{Laboratoire de Physique des Solides, Universite Paris-Sud 11, CNRS UMR 8502,
91405 Orsay, France}
\author{M.~H.~R{\"u}mmeli}
\affiliation{IFW Dresden, P.O. Box 270116, D-01171 Dresden, Germany}
\author{Th.~Pichler}
\affiliation{University of Vienna, Faculty of Physics, Strudlhofgasse 4, A-1090 Wien,
Austria}
\author{F.~Simon}
\affiliation{University of Vienna, Faculty of Physics, Strudlhofgasse 4, A-1090 Wien,
Austria}
\affiliation{Budapest University of Technology and Economics, Institute of Physics, H-1521 Budapest, Hungary}
\date{\today}

\begin{abstract}
We report $^{13}$C nuclear magnetic resonance measurements on
single wall carbon nanotube (SWCNT) bundles. The temperature dependence of
the nuclear spin-lattice relaxation rate, $1/T_{1}$, exhibits a power-law
variation, as expected for a Tomonage-Luttinger liquid (TLL). 
The observed exponent is smaller than that expected for the two band
TLL model. A departure from the power law is observed only at low $T$,
where thermal and electronic Zeeman energy merge. 
Extrapolation to zero magnetic field indicates gapless spin excitations. 
The wide $T$ range on which power-law behavior is
observed suggests that  SWCNT is so far the best realization of a one-dimensional quantum metal.

\end{abstract}


\maketitle

In a low dimensional electronic system, restricted phase space and strong
quantum fluctuations lead to emergence of unusual physical properties. A
long studied example is the one-dimensional (1-D) metal, where the
conduction electrons are confined to move along the same direction. The
standard Landau-Fermi liquid approach cannot be applied to such a system
where the high probability of electron-electron scattering prohibits a
description in terms of quasi-particles. The physical properties of a 1-D
metal in particular with a linear energy dispersion are rather 
described by the Tomonaga-Luttinger liquid (TLL) 
theory\cite{tomonaga-PTP5,luttinger-JMP4}. Characteristics of the TLL
are the occurrence of power-law divergences of the electronic response
functions at low energies. These give rise to characteristic power-law
behavior of experimental quantities as function of relevant energy scales
such as frequency or temperature \cite{giamarchi-2004}. So far
experimental studies have been limited by the difficulties in realizing the ideal
1-D metallic state. Historically, the most studied
candidates for such a realization were organic
systems, e.g. conducting polymers like polyacethylene\cite{heeger-RMP60} and
molecular conductors like TTF-TCNQ or Bechgaard salts (TMTSF)$_{2}$X
 (Ref.\cite{jerome-AP51}). 
Despite their strong anisotropy, however, these materials
still have a three dimensional electronic structure. Therefore in practice
the temperature range where the 1-D behavior can be observed is limited by
dimensional crossover effects and resulting structural instabilities.

Carbon nanotubes (CNT) appear to be a natural candidates to realize 
1-D electronic systems, because of their enormous size aspect ratio.
Experimentally, indeed, the TLL behavior was observed in the
electronic transport measurements on single wall carbon nanotube
(SWCNT) bundles \cite{bockrath-Nature397}. One might notice, however, that
the interpretation of conductivity measurements is not straightforward
because of Coulomb blockade effects\cite{postma-PRB62}. Subsequently, the
photoemission spectroscopy on bundles of metallic nanotubes also
revealed power-law behavior in the momentum distribution and the $T$
dependence of the density of states near the Fermi energy\cite{ishii-Nature426}, 
both of which are characteristics of the TLL state.

NMR experiments also provide a sensitive probe of the low-temperature
electronic state of 1-D metals, because nuclear spin-lattice relaxation rate, 
$1/T_{1}$, directly measures the local spin excitations. In addition, this
technique does not require any mechanical contacts to tubes, which can otherwise 
contaminate the electronic state. However, NMR experiment can be hardly 
performed on SWCNTs with small natural abundance of $^{13}$C ($1.1$ \%). $^{13}$%
C enrichment was done on double wall carbon nanotubes (DWCNT) using the
\textquotedblright peapod" process of synthesis. Therein, $^{13}$C NMR allows
to probe the inner tubes, which are made out of $^{13}$C enriched C$_{60}$
balls\cite{singer-PRL95}. These measurements revealed that although one
operates on a macroscopic sample involving DWCNTs with different geometries,
quasi uniform metallic properties were observed. Those were found to
display 1-D behavior with eventually a TLL power-law behavior on a limited
temperature range \cite{dora-PRL99}. Remarkably, a clear uniform gap was
detected below $T_{g}\simeq 20$ K, which should be associated with some
instability of the 1-D state.

In this Letter, we report the results of $^{13}$C NMR experiment on bundles
of SWCNTs for which optimal processes were developed in order to produce $%
^{13}$C enriched samples with minimal impurity content\cite{rummeli-JPC111}%
. The progress in isotope-engineering of SWCNTs allowed us to investigate a
large temperature range without residual magnetic catalyst contamination. In
contrast to the results on DWCNTs, we report here nearly gapless
excitations even at low temperatures on SWCNTs. The comparison between the
two systems strongly suggests that SWCNTs are a good 1-D metal and opens many
questions concerning the TLL\ state of such nanotubes.

The $^{13}$C enriched SWCNT samples used in this study were grown using the
laser ablation growth method with an isotope enriched graphite target 
($20-30$ \% $^{13}$C) 
and a mixture of non-magnetic Pt, Rh, and Re (Ref. \cite{rummeli-JPC111}). The
as-prepared samples were purified by microwave heating, oxydation in air and
washing in acids. A SWCNT content  higher than 90 \% in weight  was estimated
by optical absorption. Raman spectroscopy indicated a mean tube diameter 
of $1.6$ nm with a variance of $0.1$ nm. 
The estimated length of the tubes
appears to $600$ nm. Electron spin resonance spectroscopy
indicated the absence of either magnetic catalyst particles or residual
paramagnetic impurities. The fine powder samples were introduced into quartz
glass tubes which were sealed under $20$ mbar $^{4}$He for the NMR
measurements.

\begin{figure}[tbp]
\begin{center}
\includegraphics[width=7.5cm]{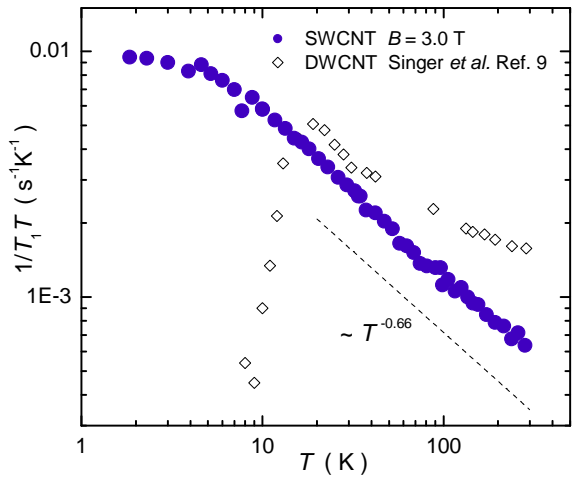}
\end{center}
\caption{
$T$ dependence of $(T_{1}T)^{-1}$ in the SWCNTs. 
For comparison similar data on the DWCNTs is shown after Ref.\cite{singer-PRL95}.
A gap is seen below $20$ K for the latter and the data
approaches a Fermi-liquid variation at room $T$. 
In SWCNTs, the low-$T$ gap is absent and 
the power-law behavior with $(T_{1}T)^{-1} \propto T^{-0.66\pm 0.01}$ 
persists up to room $T$. 
The dashed line shows the latter behavior. }
\label{fig1}
\end{figure}
%

The nuclear spin-lattice relaxation rate $1/T_{1}$ was measured with the
conventional saturation-recovery method. The  nuclear magnetization recovery 
curves do not fit the single exponential function anticipated when 
$T_{1}$ is uniform throughout the sample for a nuclear spin $1/2$. To take
the actual distribution of $T_{1}$ into account, we used a stretched
exponential fit to the relaxation curves 
\begin{equation}
\frac{M_{0}-M(t)}{M_{0}}=A\exp \left(-(t/T_{1})^{\beta } \right).  \label{stretch}
\end{equation}%
Here $M(t)$ and $M_{0}$ are the nuclear spin magnetizations at delay $t$
after saturation pulses and at equilibrium, respectively. $A$ represents
the initial saturation of the nuclear magnetization in the experimental
conditions.

The temperature dependence of $(T_{1}T)^{-1}$ in SWCNTs in a magnetic field of $3.0$ T is
displayed in Fig.~\ref{fig1} together with that on DWCNTs after Ref.~\cite{singer-PRL95}.
A power-law behavior with $(T_{1}T)^{-1} \propto T^{-0.66\pm 0.01}$ 
is observed in a large $T$ range of $6 \sim 300$ K, 
which is indicative of a TLL state maintained in a much larger $T$ range than for DWCNTs.

\begin{figure}[tbp]
\begin{center}
\includegraphics[width=7.5cm]{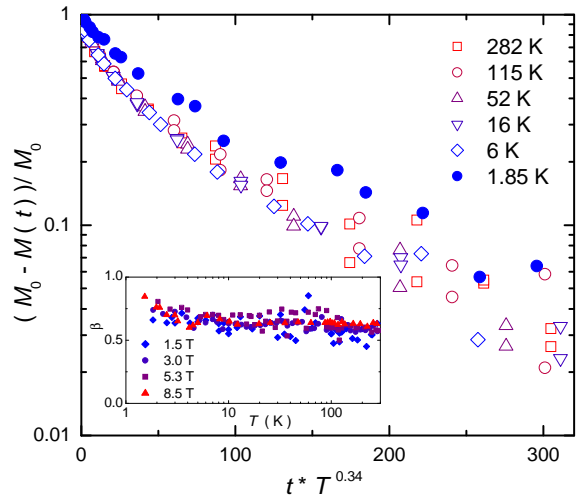}
\end{center}
\caption{
Temperature dependent magnetization recovery curves in the isotope enriched SWCNTs 
versus the product $t\ast T^{0.34}$ in $3.0$ T magnetic field.  
Data at all temperatures except the lowest $1.85$ K fall onto a single curve. 
This indicates that the distributed $T_{1}$'s follows the same $T$ dependence.
Inset: $T$ variation of $\beta $ values used for the magnetization
recovery  fits.
}
\label{fig2}
\end{figure}

The parameter $\beta $ allows to describe the distribution of relaxation
times in the sample. The fitted values of $\beta $ were found $T$
independent within experimental accuracy, 
as in the case of DWCNTs\cite{singer-PRL95}, 
with a similar average value of $\simeq 0.6$ (Fig.~\ref{fig2} inset). This rules out
the possibility that the $T_{1}$ distribution arises from distinct parts of
the sample with various electronic ground states, which would generate
highly differing $T$ dependences of $T_{1}$. It is noteworthy that, for
about $90$ \% of the nuclear magnetization, the ratio between the longest
and shortest $T_{1}$ at a fixed temperature is approximately $6$, which is
consistent with the previously reported value in both SWCNTs
\cite{fuhrer-science288} or DWCNTs\cite{singer-PRL95}. 

%

In order to prove that the exponent does not depend on the determination of $%
T_{1}$, we plot in Fig.~\ref{fig2} the relaxation curves as a function of
the product  $t\ast T^{0.34}$. The relaxation curves obtained in the
temperature range between $6$ K and room temperatures fall onto a single
curve, while that at $1.85$ K shows a substantial deviation. The scaling by
the factor $T^{0.34}$ confirms that all the distributed $T_{1}$'s are
proportional to $T^{0.34}$. 
This means that the $T$ dependence of the spin
excitations is rather homogeneous in the sample and has a truly electronic
origin, even if the actual values of $T_{1}$ are distributed. 
The relatively small distribution range as well as its independence on the type
of tubes supports that the distribution originates in both SWCNTs and DWCNTs
from the orientation of the nanotubes with respect to the applied field.


The electronic structure of SWCNTs displays two Dirac points, 
such as that of graphene. 
a linear energy dispersion found in the vicinity of Fermi energy  
gives rise to a two-band TLL behavior, 
which differs markedly from the single-band TLL for metallic 1-D chains. 
Correspondingly, the low energy excitations of a two-band TLL is 
described by two spin and two charge modes 
with four Luttinger parameters: $K_{s \pm }$ and $K_{c \pm}$.\cite{egger-EPJ3}

Assuming that the applied field is small enough 
to maintain the spin rotational invariance, 
$K_{s\pm}=1$ and only the $K_{c+}$ charge mode is affected strongly by the 
interactions\cite{giamarchi-2004,dora-PRL99}, 
its value being different from $1$. 
This allows to determine the exponents for the $(T_{1}T)^{-1}$ 
NMR relaxation rate. 
Generally, $1/T_{1}$ is expressed using the spin correlation function as 
\begin{equation}
\frac{1}{T_{1}}=\frac{A_{\mathrm{hf}}^{2}}{2\hbar ^{2}}\int d\tau \cos
\omega _{0}\tau \left\langle \frac{S_{+}(\tau )S_{-}(0)+S_{-}(\tau )S_{+}(0)%
}{2}\right\rangle ,  \label{eq:exp}
\end{equation}%
where $A_{\mathrm{hf}}$ and $\omega _{0}$ are the hyperfine coupling
constant and the NMR frequency, respectively\cite{moriya-PTP16}. The
calculations of the spin correlation function for this  1-D state gives then
the following $T$ dependence. 
\begin{equation}
\frac{1}{T_{1}T}\propto T^{\eta /2-2},\;\text{with }\;\eta =K_{c+}+3.
\end{equation}

So the $-0.66$ exponent found experimentally for $(T_{1}T)^{-1}$ would
correspond to a non realistic negative value $K_{c+}=-0.34$. This points
out that some of the assumptions made above are not valid. 
Within this two-bands approach realistic values of  $K_{c+}=0.28$ or $K_{c+}=0.18$ 
were found respectively from conductivity and photoemission spectroscopy
\cite{bockrath-Nature397,ishii-Nature426}. 
Let us point that, if the two charge modes were decoupled, the simple one-band TLL
situation, where $\eta =2K_{c}+2\ $, would be valid and yield then 
$K_{c} = 0.34$ from our data. Notice that in this single band case 
the $K_{c}$ deduced from photoemission data would be $0.28$, 
which indicates that the latter is less sensitive to the actual choice 
between one and two band models than our NMR data. 

We note on the differences observed for the DWCNTs previously. 
Therein, a similar power-law $T$ dependence was observed only between 
$20$ K and $50$ K and the NMR relaxation rate appears to deviate 
towards a Fermi liquid-like behavior around room $T$. We raise two
possibilities to account for this. 
One is a subtle change in dimensionality due to the 
inter-wall coupling between the inner and outer nanotube shells, 
which could account for the suppression of the quantum fluctuations
and stabilization of the Fermi-liquid state. 
The other possibility is an 
increase in carrier density which would screen out the Coulomb interaction
between the electrons and restore a Fermi-gas picture, where the Luttinger
parameter $K_{c}$ is equal to $1$. 
Although this might occur in electron
doped nanotubes \cite{rauf-PRL93, kramberger-PRB79}, 
it seems hard to obtain the required
densities solely by thermal population of undoped CNT.

In addition to the different high-$T$ behavior in SW- and DWCNTs, 
there is a marked difference also at low $T$. 
For DWCNTs, a gap behavior is observed below $T_g \simeq 20$ K, 
which is absent for the SWCNTs 
except for a small departure found below $T_{X} \simeq 6$ K. 

\begin{figure}[tbp]
\begin{center}
\includegraphics[width=7.5cm]{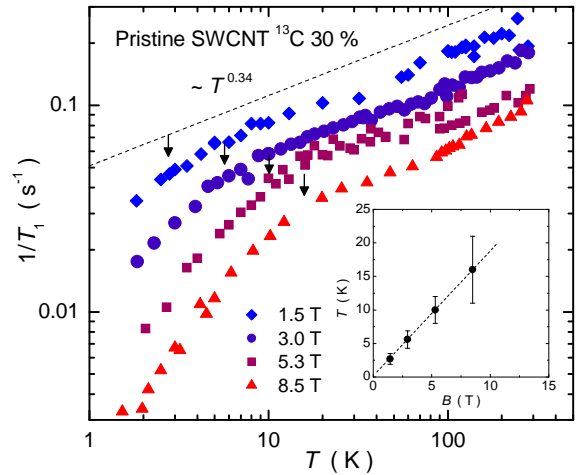}
\end{center}
\caption{$T$ dependence of $1/T_{1}$ in four different magnetic fields. The
dotted line in the figure represents the curve for 
$1/T_{1}\propto T^{0.34}$. 
The downward arrows indicate the temperatures, $T_{X}$, at which deviation occurs. 
The power law persists down to the lower $T$ in the lower fields.
Inset: Field dependence of $T_{X}$. 
The dotted line is a guide to the eyes.  
}
\label{fig3}
\end{figure}
%

The energy scale $k_{B}T_{g}$ for such gap behaviors detected on spin
excitations is found much smaller than that seen on charge excitations by
scanning tunneling microscopy (STM) in SWCNTs\cite{ouyang-science292}. As
these charge gaps were found to  decrease with increasing tube diameter, 
by analogy we tentatively assign the reduction of the gap size detected
by NMR from the DWCNT to the SWCNT to a reduction of the nanotube curvature.

\begin{figure}[tbp]
\begin{center}
\includegraphics[width=7.0cm]{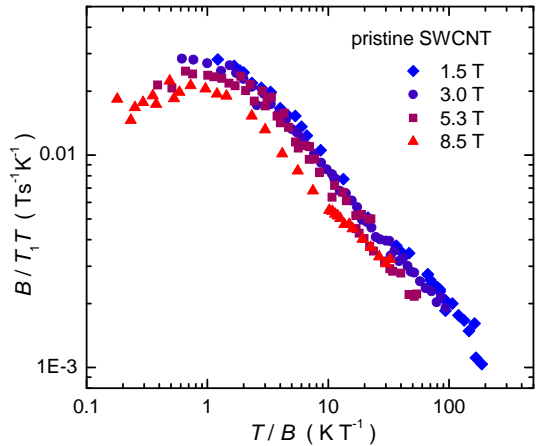}
\end{center}
\caption{
$(T_{1}T)^{-1}$ plotted versus the scaled temperature $\Gamma =T/B$. 
All the curves in four different fields fall onto a
similar function. The broad peak is observed for $\Gamma \simeq 1$. 
}
\label{fig4}
\end{figure}

To better explore the possible existence of a small gap in SWCNTs we 
studied the field dependence of $1/T_{1}$ by taking data in four
different fields from $1.5$ T up to $8.5$ T. The results are shown in 
Fig.~\ref{fig3}, where a similar high-$T$ exponent is observed for 
all fields,
although $1/T_{1}$ is found to increase slightly with decreasing field, as
was the case as well in DWCNTs\cite{singer-PRL95}. 
In contrast, while $T_{g}$ was found therein to be field independent, 
here $T_{X}$ (arrows in Fig.~\ref{fig3}) increases with the magnetic field, 
so that the power-law
variation persists down to the lowest temperatures in the lower fields. 
Since $T_{X}$ extrapolates to zero at low field, as displayed in the inset 
of Fig.~\ref{fig3}, we conclude that the spin excitations are gapless in zero
field. This also indicates that the $T_{1}$ behavior found below $T_{X}$
cannot be associated with a field independent energy scale, \textit{e.g.}
that associated with finite tube length effects. As $T_{X}$ scales
linearly with the applied field, we do find that $1/T_{1}$ scales (at least
at low $T$) with the parameter $\Gamma =T/B$ as shown in Fig.~\ref{fig4}. 
The change of slope appears then for $\Gamma \simeq 1$ (K/T) that is for 
$\mu _{B}B\simeq $ $k_{B}T$. Such a scaling is natural for independent free
spins, however it is rather surprising here as we are considering an
interacting electronic system for which the magnetic Zeeman term should only
induce a change of behavior for $\mu_{B} B$ comparable to the bandwidth.

%

When $T_{1}$ deviates from the high-$T$ variation 
$1/T_{1}\propto T^{0.34}$, the relaxation curve cannot be scaled by the same
factor as seen in Fig.~\ref{fig2}, thus the low-$T$ behavior can indeed
correspond to a distinct relaxation process. In the case of DWCNTs the
change of $\beta $ value below $T_{g}$ was attributed to 
paramagnetic impurity relaxation in the gapped state. Indeed, 
presence of magnetic impurities in clean samples limits the measurable
relaxation times $T_{1}$, yielding a minimum for $(T_{1}T)^{-1}$ at low
$T$. 

The observations done here in the absence of a gap are opposite, as $1/T_{1}$
is highly suppressed by application of a magnetic field at low $T$, 
an effect which cannot be assigned to out-of-chains impurities. One
may still wonder whether the behavior found could be governed by intrinsic
impurities embedded in the SWCNTs. It is generally
established that in correlated electron systems even non-magnetic defects,
such as chain ends or substituted non magnetic impurities reveal Curie
moments which extend on the correlation length of the pure system\cite%
{alloul-RMP81}. In the present case, due to the slow decaying correlation
functions, a moderate concentration of defects could nearly uniformly affect
the TLL state. In that case the incidence of these moments would be totally
suppressed in large fields, which would restore the behavior which is 
intrinsic to pristine SWCNTs. 
It remains an open question whether the presence of defects could explain 
the deviation of the high-$T$ exponent from the value expected for the 
two-band TLL state. 
It is also yet unclear whether the field dependence of $1/T_1$ seen 
at high $T$ in Fig.~\ref{fig4} could be attributed to a 
power-law frequency dependence of the correlation function. 

In conclusion, we performed $^{13}$C NMR measurement on SWCNT bundles.
We observed power-law behavior in the spin excitations in a wide
temperature range, strongly indicative of a TLL state. 
A departure appears in the $T$ dependence of $1/T_{1}$ below a temperature $T_{X}$, 
which depend linearly on the external magnetic fields and extrapolate 
to $T=0$ K, which evidences that the spin excitations are gapless 
down to very low $T$. 
These results indicate that SWCNTs are yet the best representation 
for 1-D metallic behavior. However, the observed exponents do not agree
with theoretical expectations for the TLL state of undoped SWCNTs. 
While we are considering a series of experiments to 
complement the present results, 
we expect that they should stimulate theoretical efforts to
account for the peculiar observations reported herein.

We would like to thank B.~D{\'{o}}ra, T.~Giamarchi and P.~Simon for
stimulating exchanges about the TLL behavior of SWCNTs, and the latter for
suggesting the possible importance of intrinsic defects. 
This work is partially supported by the projects DFG Nr. 
PI 440/3/5 and FWF Nr. P21333-N20. 
Y. Ihara
acknowledges as well financial support from JSPS postdoctoral fellowship for research abroad.

\end{document}